\documentstyle[emulateapj]{article}
\def\ltsima{$\; \buildrel < \over \sim \;$}
\def\simlt{\lower.5ex\hbox{\ltsima}}
\def\gtsima{$\; \buildrel > \over \sim \;$}
\def\simgt{\lower.5ex\hbox{\gtsima}}
\def\be{\begin{equation}}
\def\ee{\end{equation}}

\accepted{8 February 2001 }

\lefthead{Wang et al. }
\righthead{Rapid X-ray Flare in PKS 0558-504}

\begin{document}
\title{ A Rapid X-ray Flare in the Radio-loud Narrow Line Quasar PKS 0558-504}
\author{T.G. Wang\altaffilmark{1}, M. Matsuoka\altaffilmark{2}, H. Kubo\altaffilmark{3}, T. Mihara\altaffilmark{4}, H. Negoro\altaffilmark{4}}
\altaffiltext{1}{Center for Astrophysics, University of Science
\& Technology of China, Hefei, Anhui 230026, China (Email:
twang@ustc.edu.cn)} \altaffiltext{2}{Space Utilization Research
Program, NASDA, 2-1-1, Sengen, Tsukuba-city, Ibaraki 305-8505,
Japan} \altaffiltext{3}{Department of Physics,Faculty of Science,
Kyoto University, Sakyo-ku, Kyoto 606-8502, Japan}
\altaffiltext{4}{The Institute of Physical and Chemical Research
({\it RIKEN}), 2-1, Hirosawa, Wako, Saitama, 351-0198, Japan}

\begin{abstract}

We report the detection of a very short time scale flare in the radio-loud
narrow line quasars PKS 0558-504 by using ASCA. The X-ray count rates
increased and decreased by a factor of two in 33 minutes, and possibly by
40\% in as short as two minutes during this flare, confirming that such flare
event does occur in this object with imaging detectors. The implied largest
rate of
change in luminosity in 0.8-10 keV alone, $dL/dt \simeq (1.8\pm 0.4)$ 10$^{42}$
ergs s$^{-2}$, is several times
higher than the limit sets for the isotropic emitting plasma around a Kerr
black hole.
Either emission from a relativistic boosting jet or a magnetic heated corona
may explain such high radiative efficiency.
Magnetic field with a strength of at least a few $10^4$ Gauss is required
 in the latter case.  The spectrum during this flare is significant harder
than the average one. Three radio loud narrow line active galactic nuclei (AGN) 
possess considerably smaller black holes than in radio loud AGN in bright 
quasar survey, indicating
 they are still in a phase of the rapid growth of the black hole by accretion.

\end{abstract}

\section{Introduction}

Most Active Galactic Nuclei (AGN) show some variations in fluxes. The
timescale and amplitude depends strongly on the wavelength observed:
particularly, the amplitude is larger and time scale shorter in the X-ray band
than in the optical one. Large amplitude variations over a time scale
from minutes to hours have been reported for Seyfert galaxies and a
particular beamed class of AGN named blazars (e.g., Barr \& Mushotzky
1986, McHardy 1989, Catanese \& Sambrana 2000). A physical causality
argument leads to either very small size of the emission region and/or
relativistic beaming in these objects. In fact, combining with the large
energy output in the X-ray band, this is considered one strong evidence
for their extremely high radiative efficiency, which leads to the general
picture of massive black hole accretion for AGN (e.g., Fabian
1979, Rees 1984).

Among various types of AGN, only Seyfert galaxies and BL Lac
objects are well studied for short time scale X-ray variability.
This is partly due to their brightness in the X-ray band. Nandra
et al. (1997) presented a systematic analysis the variability of
X-ray flux for a sample of broad lined Seyfert galaxies observed
by ASCA and found that the X-ray variability amplitude is well
anti-correlated with the luminosity. Leighly (1999) demonstrated
that a subclass of Seyfert galaxies called Narrow Line Seyfert 1
galaxies (NLS1s\footnote{We do not distinguish NLS1 and narrow line 
quasar, and call them simply NLS1}, Osterbrock \& Page 1985, Goodrich 
1989) also follows an anti-correlation of amplitudes with luminosities, but
for a given X-ray luminosity, NLS1s shows much larger variability
amplitudes. This finding confirms the suspicion, based on the
studies of individual objects, that NLS1s are more variable than
normal Seyfert galaxies. Example of large variations can be found
in literature (e.g., Boller, Brandt \& Fink 1996, Forster \&
Halpern 1996, Otani, Kii \& Miya 1996).

PKS 0558-504 (z=0.137, $m_B$=14.97) is one of the rare radio-loud
NLS1 type objects. Remillard et al. (1992) reported the detection
of a rapid flare, in which the X-ray flux increased by 67\% in 3
minutes, using the Ginga Satellite and suggested that the X-ray
emission is enhanced by relativistic beaming. However, lack of
imaging capacity of the Ginga makes this result somewhat
questionable, as the flare might be due to a neighboring source.
Recently, Gliozzi et al. (2000) observed this object with the
ROSAT HRI in the soft X-ray band, they found also strong
persistent variations in the soft X-ray band. But the $\Delta
L/\Delta t$ is considerably smaller than that reported by
Remillard et al. (1992). In this paper, we report the detection
of a rapid energetic flare in this object with imaging
detectors.

\section{Observation and Data Reduction}

PKS 0558-504 was observed with ASCA (Tanaka et al. 1994) on January 31,
2000, with an effective
exposure time of 37.4 ksec. The Solid-state Imaging Spectrometer (SIS) was
operated in mixing faint and Bright 1-CCD mode, and the Gas Imaging
Spectrometer (GIS) on pulse height mode. For the SIS, the faint mode data
were converted into bright ones and combined with the bright data. The
data reduction was performed in the standard way by using the FTOOLs
(v4.2). Hot and flickering pixels were removed from the SIS data. To check
the reliability of the scientific results, we have chosen both the strict
data screening criteria and standard ones (see The ASCA Data Reduction
Guide, version 2). The two yields consistent results, particularly on the
flare profile. Since the standard one yields slightly better statistics in
the X-ray spectrum, we will present it below. After standard screening,
the net exposure times are 37.4 ks and 25.9 ksec for the GIS and SIS
detectors, respectively.

The source counts were extracted from a circular region of 3.5 and 5.0
arcmin radius for SIS and GIS, respectively. The background counts were
estimated from the off-source region at the same off-axis angle with same
area for each of GIS detectors, and from the source subtracted region of
the same chip of CCD for each of the SIS detectors. The background accounts for
about 5 percent of total counts both for the SIS and GIS. The average
net source count rates, after correcting for the background, are
0.470$\pm$0.003, 0.598$\pm$0.004, 1.009$\pm$0.007 and 0.820$\pm$0.005
cts~sec$^{-1}$ for GIS2, GIS3, SIS0 and SIS1, respectively.

Light curves were extracted for the source and background for each detector.
For both GIS detectors, the background count rate is nearly constant
during the observation, therefore, an average count rate is used to
estimate the background level. For the SIS detectors, the extracted
background rate seems correlated with the source due to the contamination
of the AGN. Therefore, it is only an upper limit to the true background,
and is used only for spectral analysis.
Since smaller extraction radii in the real sky are used for SIS detectors,
the fraction of background light, thus, is estimated to be smaller than
in the GIS case. We
will ignore the background contribution to the SIS count-rate during the
light curve analysis.

The X-ray spectra were rebined to at least 25 counts per each bin. For the
SIS spectrum, the response matrices appropriate for the date of
the observation (thus accounting for decline of the energy resolution as
a function of time) were made using the script {\em sisrmg}. For the GIS
spectrum, the 1994 response matrices ({\em gis2v4\_0.rmf} and {\em
gis3v4\_0.rmf} ) were adopted. Ancillary response files were made for each
detector using {\em ascaarf}. The ASCA data preparation and the spectral
analysis were performed using version 1.4 of the XSELECT package and
version 10.01 of XSPEC.

\section {Spectral And Temporal Analysis}

\subsection{The Spectral Properties}

Since the efficiency of SIS detectors has decreased due to the radiation
damage and the current calibration files does not account for this,
SIS spectra only above 0.8 keV are used. The GIS spectra below 0.8 keV
are not well calibrated and will be not used in the spectral fit. The
X-ray spectra in the full ASCA band cannot be adequately
described by an absorbed power-law with a $\chi^2/d.o.f=1058/924$, which
is accepted at a probability of only 1~10$^{-3}$. There are systematic
deviations at low energies. In addition the fitted column density
(2.4$\pm 1.2 $~10$^{20}$ cm$^{-2}$)\footnote{Uncertainties are given at
90\% confidence level for one interesting parameter for all the spectral
parameters given in this section} is significantly lower than the
galactic value (4.5~10$^{20}$~cm$^{-2}$).

In order to see if this is due
to soft X-ray excess, which was noticed in the Ginga spectrum (Remillard et
al. 1992), we initially fitted the spectrum above 2.0 keV. Single
power-law with the Galactic absorption provides a good fit to the joint
GIS and SIS spectra ($\chi^2/d.o.f$=631/636). The best fitted photon index
($\Gamma=2.20_{-0.04}^{+0.02}$) is slightly flatter than the previous fit
($\Gamma=2.25_{-0.03}^{+0.02}$). This spectral index matches well the one
obtained by Leighly et al. (1999) for 1997 observation, suggesting no
significant variation in the spectral index between the two observations.
Extrapolating this fit to low energies shows excesses in the soft X-ray
band (Fig. 1), particularly below 1.2 keV. The excess is larger in the GIS
spectra than in the SIS ones, and also larger in the SIS0 than in the SIS1,
possibly indicating the decreasing efficiency of SIS already evident in the
energies just below 1.2 keV. If the soft excesses are modeled with black body
emission, the best fit yields
a kT=0.22$\pm$0.02 keV and a normalization
$3.4_{-1.0}^{+1.3}\times$10$^{-5}$. The latter corresponds to a flux in
0.8-2 keV flux of 2.1$\times$10$^{-12}$ ergs~sec$^{-1}$~cm$^{-2}$.
Though this fit is better than the single power-law fit by $\Delta
\chi^2$=31, however, it is still statistically acceptable only at 1\%
probability ($\chi^2/d.o.f.$=1027/923). After carefully examining
the residuals, we find
that counts in the SIS detector is significantly lower than these of
GIS at energy below 1.15 keV. This most likely is due to the degeneration
of the SIS sensitivity at low energies. In fact, the fit is acceptable
(926/894, $P_r$=0.22) when the SIS data below 1.2 keV were ignored. This fit
yields an kT=0.13$_{-0.02}^{+0.03}$ keV and a normalization
1.50$_{-0.53}^{+1.10}$ 10$^{-4}$. The photon index for this fit is
2.22$_{-0.03}^{+0.02}$, similar to that derived for fitting 2-10 keV
spectrum.

Recently, O'Brien et al. (2000) found that the soft excess extends up to 3
keV from their much higher quality XMM spectrum and can be modeled as multiple
temperature blackbody emission. The spectral slope of the power law component
is 0.9,  fully consistent with those found in other Seyfert 1 galaxies.
The steep hard X-ray component found in the ASCA spectrum could be due to
the contamination of the spectroscopically unresolved soft X-ray excess in the
ASCA 2-10 keV band.

No iron K line is detectable with the ASCA data. An upper limit of equivalent
width for a narrow Gaussian line at 5.5 or 5.7 keV (6.4 and 6.7 keV in the
source rest frame) is 40 eV, consistent with XMM result (O'Brien et al 2000).

\subsection{Temporal Properties}

Fig. 2a shows the combined GIS and SIS light curves for the count rates in
0.8-10 keV and 0.6-10 keV bands, respectively. A large energetic flare
started shortly after the observation.
The count rate increased from 0.54 $\pm$ 0.02 cts/s to 1.10$\pm$ 0.04
cts/s in about 2000 sec and then decreased to 0.50$\pm$ 0.02 cts/s in
about 3500 sec. Since there are observation gaps in between, the real
variations might be even faster.

We have examined possible contamination sources. The background
count rates are stable and at only a few percent level of the
source count rates during the observation. Concerning the impact
of particle background, we have applied strict screening criteria
to the GIS data (refer to The ASCA Data Reduction Guide, version
2), the flare structure remained. The source position measured in
the ASCA GIS image is 05 59 41.1: -50 27 24.1 (equinox 2000),
which has an offset of $\Delta\alpha=1.58'$ and
$\Delta\delta=0.39'$ from the NED position of PKS 0558-504.
However, after correcting for a temperature dependent deviation
of the attitude solution (Gotthelf et al. 2000), the ASCA position
deviates from the NED one of about 13$''$, which is within the
90\% error circle (24$''$) of the GIS. We also noted that the
image extracted for the events collected during the flare phase
has a position 09 59 40.4 -50 27 16.8, which is only 11$''$ away
from the position of the source in the total observation for the
GIS. This analysis sets that any confusing source should be
within 20$''$ of the PKS 0558-504. The analysis of the SIS data
gives a similar conclusion, but detailed number is not given here
since our main analysis below concentrates on the GIS curve.

The SIS curve is similar to the GIS one. The flare is also clearly
present in the SIS data. Due to different screening criteria
applied to the SIS and GIS, the SIS data just before the peak of
flare are not available. Thus, the flare profile is less well
defined than in the GIS curve.

The flare is shown with increased time resolution (GIS 64-s bins) in Fig. 2b.
Clearly, the source shows variability on short time scales of 10$^2$s.
A $\chi^2$
test on the constancy of count-rate for the second group of the data in Fig
2b gives a $\chi^2/d.o.f$=45.6/21, which is at a probability $P_r$=0.002 by
chance.
The pre-flare has a mean count rate of 0.53$\pm$0.02 cts/s for an
interval of 650 seconds.
Unfortunately, there is a 1000 sec gap
just before the on-set of the flare. Nevertheless, the flare rose very
fast. The count rate increased from 0.82$\pm$0.05 cts/s (average over three
bins) to the peak count rate 1.14$\pm$0.05 cts/s (three bins average) in
less than 128 sec. The statistical significance of this change is about
4.5 $\sigma$. This gives a rate of change in the count-rate $\Delta CR/\Delta t
\simeq (2.5\pm 0.6) 10^{-3}$ cts~s$^{-2}$. A more subjective estimation of the
rate by linear fit to the rising part of the curve (first 7 points in the
second group of data, see Fig 2b). This gives a rate $\Delta CR/\Delta t \simeq
(1.14\pm 0.28)$ cts~s$^{-2}$, a factor of two lower than the one from direct
visual inspection. We will quote the latter more conserved number in later
discussion.
The source was then dimming with possible flicker till the
end of this orbit.  The presence of the second peak probably is also real
since it appears in the SIS curve as well. The flux returns to the
pre-flare level at the beginning of the next orbit observation.

We define two bands based on the spectral range of the soft X-ray excess
(see \S 3.1). The energy ranges are 0.8-2.0, 2.0-10 keV for soft and
hard bands for GIS data. The soft band extends to 0.5 keV for
the SIS data since there is a substantial fraction of counts in the
energy range 0.5-0.8 keV.  The light curves for these two bands were
extracted in order to examine possible spectral variations during the
flare. The flare is seen in  both bands. However, it shows larger
amplitude and decays faster in the hard band than in the soft one. The ratios
of the count rate for the flare peak to the pre-flare are 1.70$\pm$0.13
and 2.13$\pm$0.17 for the
soft and hard bands, respectively. There is no significant decrease
in the count rate of the soft band during the flare orbit ($\chi^2/d.o.f$=
3.4/3, $P_r>0.3$ for
a constant count rate in the last 1024 seconds during the flare orbit), in
contrast to the clear rising and fading in the curves of hard band
($\chi^2/d.o.f$=15.0/3, $P_r\simeq 0.002$ for the same $\chi^2$ test) (Fig 3).

We examine whether there is a correlation between the hardness ratio
and the total count-rate and whether the flare follows the same correlation.
The hardness ratio is the ratio of count rates in the 2-10keV band and in the
0.8-2.0 keV band. In order to achieve a reasonable S/N ratio for each bin,
we use a 1024 sec binning light curves. The hardness ratio is weakly
anti-correlated with the total count-rate for normal variations. A Spearman rank
correlation analysis gives a correlation coefficient of r=-0.371 (n=62) for
which
the probability for null hypothesis is $P_r$=0.3\% for the GIS data (we have
ignored these data points with uncertainty in the ratio larger than 0.15).
In figure 4, we add the hardness ratios
for the flare phase, which is binned with 512 seconds bining.
The flare is distinguished itself from others for having large hardness
ratio and large count-rate (Fig. 4).  In fact, the
first three data points of the flare phase have hardness ratio among the
largest. And the last data point of the flare returns to the average hardness
ratio. If the spectrum during the flare is a power-law, the
power-law indices is about 1.9 for the first three points of the flare and
2.2 for the last point of the flare (Fig. 4). It is not clear that if the
decreasing hardness ratio indicates a time delay between the two bands.
Cross correlation analysis does not found any definite delays between
the soft X-rays and the hard X-rays.

Since the first three data points during the flare imply a
spectral index of 1.9 (see Fig. 4), which is significantly harder
than the mean spectral index in the 2-10 keV band,
 the change in the hardness ratio cannot be explained with the change of
relative contribution of the soft excess alone. The spectrum in
the 2-10 keV band must also have changed. However, if the hard
X-ray spectrum is as flat as $\Gamma=1.9$, suggested by the XMM
observation (O'Brien et al. 2000), then during the large flare,
the X-ray emission is dominated by the hard power-law.

\section{Discussion}

We observed a rapid flare in the PKS 0558-504, during which the X-ray
count rate increased by a factor of nearly two in 33 minutes, and possibly 40\% in
as short as two minutes. This result independently confirms the existence
of a rapid flare in this object obtained by Ginga (Remillard et al., 1992)
and further suggests that the flare is repetitive. ROSAT HRI observation
also found that the object is highly variable in the soft X-ray band, but
no such clearly rapid flare event was seen (Gliozzi et al. 2000). This
might be due to the relatively flat spectrum during the flare phase (a much
lower amplitude in the soft X-ray band) and/or to the low photon collecting
 area of the ROSAT HRI or it does not happen to catch a flare. Neither the ASCA
observation in 1999 nor the XMM observation in 2000 (Gliozzi et
al 2001) detected a rapid flare. The total exposure time, by
summing up those of Ginga, ASCA and XMM observations, is about
150 ks, in which rapid flares were detected twice.
 This suggests that rapid flares occur not in-frequently in this object.

The fastest variation during the rise of flare phase suggests a
rate of change in the count rate $\Delta CR/\Delta t = (1.14\pm
0.28)~10^{-3}$ cts~s$^{-2}$ in a linear fit. Assuming the X-ray
spectrum can be described by a power-law, the flare has a
spectrum with a photon index of $\simeq$ 1.9 from its hardness
ratio, the rate of change in the count-rate yields a $\Delta L/
\Delta t = (1.8\pm 0.4) 10^{42}$ ergs s$^{-2}$ in the 0.8-10 keV
band (assuming $H_0$ = 75 and $q_0 = 0.5$) if the X-ray emits
isotropically. Note this value is similar to the one obtained
during the Ginga observation. Since the flare spectrum must be
not limited in the 0.8 to 10 keV band, the actual $\Delta L/
\Delta t$ should be larger. If the hard X-ray spectrum extends to
energy as high as 100 keV, then the flare luminosity will be a
factor of 2 higher.  Notice that there is a limit on $\Delta
L/\Delta t\leq 2~ 10^{42}\eta$ ergs~sec$^{-2}$, where $\eta$ is
the efficiency of converting matter to radiation, for spherically
distributed plasma whose opacity is dominated by Thomson
scattering (e.g., Guilbert, Fabian, \& Rees 1993). The $\Delta L/
\Delta t$ given above yields an efficiency of $\eta= 0.9\pm 0.2$,
which exceeds the limit for accretion even onto a Kerr hole. This
was interpreted as an evidence for the relativistic beaming in
this object by Remillard et al. (1991).

\subsection{A Flare from Relativistic Jets ?}

Since PKS 0558-504 is a radio-loud quasar, we will first examine
the possibility that the flare is produced in relativistic radio
jets. There are evidences that the X-ray is dominated by the
emission from relativistic jets in radio loud quasars (e.g.,
Worral \& Wilkes 1990). Relativistic beaming is expected
naturally in this case if the jet beams towards us. We noticed
that the broad band optical/UV to X-ray spectrum is flat with a
spectral index $\alpha=(1.1-1.3)$, in the range for radio-loud
quasars (Brinkmann, Yuan \& Siebert 1997). Lack of a detectable Fe
K$\alpha$ line is also consistent with the X-ray observations of
other radio loud quasars. However, the X-ray spectrum of PKS
0558-504 in the 2-10 keV band is much steeper than that of a
typical radio loud quasar. It resembles those of High energy
peaked BL Lac (HBL) objects, in which the X-ray spectrum is
believed to be the tail of the synchrotron emission of the
relativistic jets (e.g., Kubo et al. 1998). Furthermore, we
notice that HBL objects also show a loop structure on the plane of
hardness ratio versus count rate during flares, possibly due to
the synchrotron cooling of electrons (e.g., Kataoka et al. 2000).
Since the jet in a BL Lac object is thought to beam towards the
observer, this meets the requirement of relativistic boosting.
Even for BL Lac objects, such fast ($\le 10^3$s) flares were only
reported in a few cases (Feigelson et al. 1986; Catanese \&
Sambrana 2000)

But PKS 0558-504 is not BL Lac object. It has prominent emission
lines, instead of weak or non-detectable emission lines for a
BL-Lac object. In addition, PKS 0558-504 shows a steep radio
spectrum with a 2.7 GHz to 5 GHz spectral index of 0.88 (Wright
\& Otrupcek 1990, Gregory, Varasour, Scott, \& Condon 1994, and
Wright, Friffith, Burker \& Ekers 1994), in contrary to the flat
radio spectra of BL-Lac objects. And finally, the fact that the
continuum of PKS 0558-504 is dominated by a big blue bump
extending to the soft X-ray band suggests that the non-thermal
emission from jets unlikely overwhelms the nucleus emission  by a
large factor even in X-ray band (O'Brien et al. 2000).

This does not rule out the possibility that the X-ray flare is
induced by a relativistic jet while the emission is from the
nucleus in the normal state. The fact that the X-ray spectrum
during the flare phase tends to be flatter than the average
spectrum is consistent with it being from a distinct component.
If the nucleus component has a flux similar to the average value,
half of the count rate during the flare must come from the
nucleus component since count rate during the flare is a factor
two of the normal state. Then the flare component could be as
flat as a power-law with an index 1.6.

\subsection{A Magnetic Flare?}

Given that there is no evidence of associating the flare with
radio jets in PKS 0558-504, we consider the possibility that it
arises from the nucleus. Rapid variations in the X-ray band are
common even in the radio quiet NLS1s, where no relativistic radio
jet has been observed. Some extreme variations in objects such as
PHL 1092, IRAS 13224-3809, also suggest a radiative efficiency
larger than those setting by accretion matter onto Schwarzschild
black holes (e.g., Forster \& Halpern 1996, Boller, Brandt \&
Fink 1996). As discussed by Guilbert, Fabian \& Rees (1983) that
either non-spherical geometry or continuous acceleration of
electrons can generate a large effective radiative efficiency.
Particularly interesting is a magnetic corona model, in which
particles in the corona are continuously heated up by magnetic
reconnection and up-scatter the soft photons from the accretion
disk into X-rays. Since the magnetic field
 does not contribute to the scattering opacity, this will raise the effective
efficiency.  If the energy is pre-stored in the magnetic field
co-region with the emission plasma, the effective efficiency can
be as large as $\eta \le B^2/(8\pi \rho_0 c^2) $, where $B$ is
the strength of magnetic field, $\rho_0$ the mass density of the
corona region. When the magnetic energy density approaches to
that of the rest mass, the $\eta$ is order of 1. In this model,
the X-ray spectral index is a function of the particle
temperature and the optical depth of the corona. The temperature
is determined by the balance of the heating and cooling rate since
the Compton cooling is very effective for hot electrons in NLS1s
as estimated below.

The electron energy loss rate is

\begin{equation}
 \frac{dE}{dt}=n_e\int\frac{U_\nu}{h\nu} c\sigma_T
\frac{4kT-h\nu}{m_ec^2}
h\nu d\nu
\end{equation}

where $U_\nu$ is the specific radiation energy density at frequency $\nu$,
$\sigma_T$ the Thomson cross section, $m_e$ the rest mass of an electron,
and $T$ is the electron temperature.  If the average photon energy is much
smaller than the thermal energy of hot electrons, then we can ignore the
process of transferring photon energy to an electron. One can easily get
a cooling time scale for hot electrons:

\begin{equation}
\tau_c \simeq \frac{3\pi}{2} \frac{m_ec^2}{\sigma_T} \frac{r^2}{L} =
58 L_{45}^{-1}r_{14}^2 sec
\end{equation}

Where {\it L} is the luminosity in the UV and soft X-ray bands, {\it r} the
size of the continuum emission region, $L_{45}=L/(10^{45}$ ergs/s) and
$r_{14}=r/(10^{14}$ cm).

For typical luminosity and size of emission region, this time-scale is rather
short
in comparison with the typical flux variation time scale. Therefore,
if the hard X-ray emission is produced by the Compton scattering process,
the fading time (order of $10^3$ sec) seems irrelevant to the electron cooling
process, but more likely due to variations of electron heating rate, e.g., magnetic
reconnection rate or even the coupling time of electrons with ions.

If a flare is energized by magnetic
reconnection, then the total energy stored in the magnetic field
$(B^2/8\pi)l^3$ must be not less than the total energy emitted during the flare
($\Delta L\Delta t$), where {\em l} is the scale of the magnetic field, $\Delta
L$ and $\Delta t$ are the luminosity and the duration of the flare. When an
anomalous resistivity is introduced, the magnetic reconnection takes place
and it spreads at a speed of Alfv\'en velocity,
$v_A=(B^2/4\pi\rho)^{1/2}$.
The global magnetic field dissipation time should be not shorter
than $l/v_A$.  Notice that converting a significant power into
X-rays needs a Thomson optical depth $\tau_T$ of orders of 1.
Putting all these together, one yields the time scale of
variability:

\begin{equation}
\Delta t \geq 4~10^3 \Delta L_{45}^{1/5} B_4^{-8/5} \tau_T^{3/5} sec
\end{equation}

where $B_4=B/10^4$ Gauss, $\Delta L_{45}=\Delta L/(10^{45}$ ergs/s).
$\Delta t$ is sensitive to magnetic field strength, but only weakly
depends on the luminosity. For the fastest variation during the rising phase
of the flare, $\Delta L=5\times 10^{44}$ ergs/s in $\Delta t= 280$ sec
(using the linear fit result), one
yields  $B\ge 5\times 10^4 \tau_T^{3/8}$ Gauss for the flare region. If part of
the energy dissipated via magnetic reconnection is converted into kinetic
energy, causing bulk motion of the flaring material (Beloborodov 1999). The
bulk motion would boost the apparent variability if it is towards the
observer. However, the flare material is only mildly relativistic, as
estimated by Beloborodov(1999), this would not seriously affect the magnetic
field given above.

The strength of the magnetic field in an accretion disk is in
principle limited by equipartition with the disk pressure.
Stronger magnetic field will rise buoyantly from the accretion
disk, leading to a magnetically  confined corona (Galeev, Rosner
\& Vaiana 1979). Mineshige et al. (2000) argued that the magnetic
field is large in the NLS1s due to large pressure caused by the
trapped photons in the inner region (radiation pressure
dominated) of a slim disk. By requiring $P_{mag}\le
P_{disc}\simeq P_{rad}=aT^3/3$, one can estimate the temperature
of the disk of the region that produced the flare.
 This gives a $kT_{disk}\ge 38 eV$, which is somewhat lower than
that of the lowest temperature component in O'Brien et al.'s
multiple black body model derived from the broad band XMM spectrum.
 This perhaps explains why a rapid energetic flare could only observed
in NLS1s with extremely steep soft X-ray.


\subsection{The Black Hole Masses in Radio Loud NLS1s}

There is growing evidence for more massive black holes (BH) in
radio-loud quasars than in radio quiet ones. By assuming the line
emitting gas is virialized, Laor (2000) derived BH masses of the
radio loud QSOs above $10^9$ M$_\odot$, significantly larger than
in radio quiet QSOs. This result is in line with the recent
identified correlation between the mass of the central BH and
that of the bulge (Magorrian et al. 1998, Gebhard et al. 2000,
Ferrarese \& Merritt 2000), together with the fact that the radio
loud objects are hosted by elliptical galaxies. It is not
understood how the jet formation is related to the mass of the
central BH in AGN, the galactic BH binaries also show
super-luminal radio jets but their typical mass is around
10M$_\odot$\footnote{They are formally radio-quiet according to
the definition of radio loudness $R=log (f_{5GHz}/f_{opt})$ in
AGN.}. McLure et al. (1999) found that both luminous radio-loud
and radio quiet quasars reside in elliptical galaxies, suggested
massive BH for both type QSOs.

On the other hand, it was suggested that the formation of jets might be
related to the low accretion rate, such as in advection dominated accretion
flow (e.g, Rees, Begelman, Blandford \& Phinney 1982, Blandford \& Begelman 1999)
 and/or with the spin of the BH ( Blandford \& Znajek 1977). It is worthy
to mention that the radio emission from the BH X-ray binaries also
appear to be strong in the low state (Fender 2000), while it is suppressed in
the high state.

However, evidence shows that the PKS 0558-504 possesses a low mass BH
and high accretion rate.
The H$\beta$ width of this object is around 1250 km s$^{-1}$
(Corbin 1997) and the optical continuum luminosity $\nu
L\nu(5100\AA) \simeq 2.2~10^{45}$ erg s$^{-1}$ from the V
magnitude and corrected for the Galactic reddening (Corbin \&
Smith 2000). By adopting the empirical Broad Line Region (BLR)
size  versus optical luminosity relation, $R_{BLR}\simeq 18.65
[\lambda L\lambda(5100\AA)/10^{44} {\rm ergs~s}^{-1}]^{0.7}$ (
Kaspi et al. 2000), and further assuming that the broad line
region is virialized, we can derive a mass of central BH of
$4.5\times 10^7$ solar mass. With this BH mass, the object is
emitting at super-Eddington luminosity. Note the mass of the
central BH is far less than those seen in the radio loud quasars
in the PG sample (Laor 2000), which was inferred with the same
method. We wish to point that the V-magnitude of PKS 0558-504 is
14.97 and its B-V, U-B values similar to those of 3C 273. If it
were in the north sky, it would be selected as a PG QSO as well.

There are a number of uncertainties in the above estimation of
the mass. We have used the empirical $R_{BLR}-L_{opt}$ relation.
The BLR is photo-ionized by far/extreme UV photons, the size may
scale more adequately with the ionizing continuum than with the
optical luminosity. Since PKS 0558-504 shows very big blue bump as
detected by XMM in optical-UV and broad band X-ray (O'Brien et
al. 2000), it has more ionizing flux than a typical QSO with
similar optical luminosity, thus a larger BLR. To give the order
of magnitude of this correction, we estimate an optical to UV
spectral slope $\alpha\simeq 0.1$, which is 0.4 lower than
typical QSO value. If this spectrum extends to the Lyman limit,
there is a factor of 2 more flux at this 912\AA~ than for typical
QSO spectrum. This will only increases the mass of BH  by a
factor of 1.5, which is still far less than the masses of BHs in
radio loud PG QSOs. Since by this way, the increase of ionizing
photon flux also raises the bolometric luminosity, this would not
lower the fraction accretion rate. Another concern is that the
BLR is in a flat structure, so that the velocity depends strongly
on the inclination of the system.  There is evidence for the
anisotropy of BLR velocity in radio loud AGNs, and the H$\beta$
line width is found to be anti-correlated with the core dominance
of the radio source. There is no such radio data for the PKS
0558-504, so it remains possible that we might see a flat BLR
from top in this object and the mass of central BH was
under-estimated.  However, the core dominated radio source
usually display flat spectrum, instead of the steep radio
spectrum in these objects (see \S 4.1). High resolution radio observation, 
however, is necessary to address this.

The other two radio loud NLS1s (RGB J0044+193 and RX
J0134.2-4258) have even lower BH masses and higher accretion
rates. Using the same method, we find that the BH masses are
$1.6\times 10^7$ and $10^7$ M$_\odot$ for RGB J0044+193 and RX
J0134.2-4258, respectively. These two objects show the same
characteristics in the SED as PKS 0558-504: a very big blue bump
indicated by its flat optical to UV continuum and huge soft X-ray
excess, steep radio spectrum, and weak iron K line (Siebert et
al. 1999, Grupe et al. 2000). The mass of RGB J0044+193 can be an
order of magnitude larger if the BLR size scales with the
luminosity of ionizing continuum instead of the luminosity at
5100\AA~ and if its extremely steep continuum in the optical
spectrum ($\alpha_{opt} \simeq -3.1$, Siebert et al. 1999)
extends into far UV. However, we notice that the equivalent width
of H$\beta$ is normal in this object, suggesting such kind of
correction is not adequate.

If above mass estimation is correct, the existence of radio-loud
NLS1s suggests that neither large mass of the BH nor a low
accretion rate is necessary for the formation of a powerful radio
jet. So what else factor is relevant, spin of a BH or the
environment of the galactic nuclei? If a jet is powered by the
spin energy of the BH through Blandford \& Znajek (1977)
mechanism, both rapid spin of the BH and a strong magnetic field
are required for the formation of powerful radio jets. Formation
of a rapidly spinning BH in NLS1s can be found in various
schemes: a massive hole is formed through collapsing of a
rotating gas clouds, through accreting material by a seeded small
hole and by merging of small holes. In particular, the hole spins
up very fast in NLS1s as they are thought to accrete at close to
or even supper-Eddington rate. It can easily reach the equilibrium
between the spin-up by accretion and the spin-down by the
Blandford-Znajek mechanism (Moderski, Sikora \& Lasota 1998)
unless the NLS1s phase is extremely short (less than a few $10^7$
years). Strong magnetic field can be either created by the dynamo
in the accretion disk or amplified by compressing the convected
magnetic field frozen in the accreted material. It remains
unexpected that radio loud NLS1s are so rare.

Alternatively, it was suggested that the observed radio jets are related
not only with the center engine but also with the environment of the nuclei
that provide the confinement of the jets. If this is indeed the case, the host
galaxies of the RL NLS1s should be also a massive elliptical galaxies. We
speculate that they deviate from the relation between mass of BH and that of
bulge because the BH is still in the rapid growth phase,
We have no data on the bulge masses of these three NLS1s. Nelson
\& Whittle (1996) suggest that [OIII] $\lambda 5007$ width is a
good indicator of stellar velocity dispersion in elliptical
galaxies and spiral bulges.  The [OIII] widths are 670 and 750
km~s$^{-1}$ for RX J0134.2-4258 and RGB J0044+193 (Siebert et al
1999; Grupe et al. 2000), respectively, which correspond to
stellar velocity dispersions of 285 and 319 km s$^{-1}$,
indicating of massive spheroidal component in both galaxies. This
would give BH masses of about 4-6 $10^8$ M$_\odot$ in these two
objects if they were followed BH mass and stellar velocity
dispersion relation (Fig. 1 of Nelson 2000). These masses are one
order of magnitude larger than the BH masses estimated from the
BLR-kinematics method, consistent with our guess. The [OIII] is
extremely weak in PKS 0558-504 and its has not been given by
Corbin (1997), but seems also relative broad. However, a direct
measurement of the stellar velocity dispersion in those objects
is needed before any firm conclusion can be drawn.

To summarize, we find that the rapid flare observed in PKS 0558-504
requires an effective radiative efficiency of close to one. This can be
explained either by associating the flare with the relativistic radio jets
or magnetic heated corona above an accretion disk.  Future simultaneous 
monitor of this object in radio and X-ray bands would be crucial in 
discrimination the two possibility. We showed that a magnetic
field of strength of at least a few $10^4$ Gauss is required in the latter
case.  Future observation with a large photon collecting areas detector, such
as MAXI program, will allow un-interrupt monitoring the detail spectral
evolution of the flare, yielding stringent constraints on the models. We
found that the masses in three RL NLS1s are much lower than that for radio loud
quasars, which may suggest that the black holes in narrow line NLS1 are still
in the phase of rapid growth.

\acknowledgments{The authors thank the ASCA team for successfully
carrying out this observation. TW wishes to thank Youjun Lu for
many useful comments on an early version of this draft. TW
acknowledges the financial support of Chinese NSF through
NSF-19925313 and of Chinese Science and Technology Ministry. This research 
has made use of the NASA/IPAC Extragalactic Database (NED) which is operated 
by the Jet Propulsion Laboratory, California Institute of Technology, under 
contract with NASA.}

\newpage
\begin{figure}
\figurenum{1}
\epsscale{1.00}
\plotfiddle{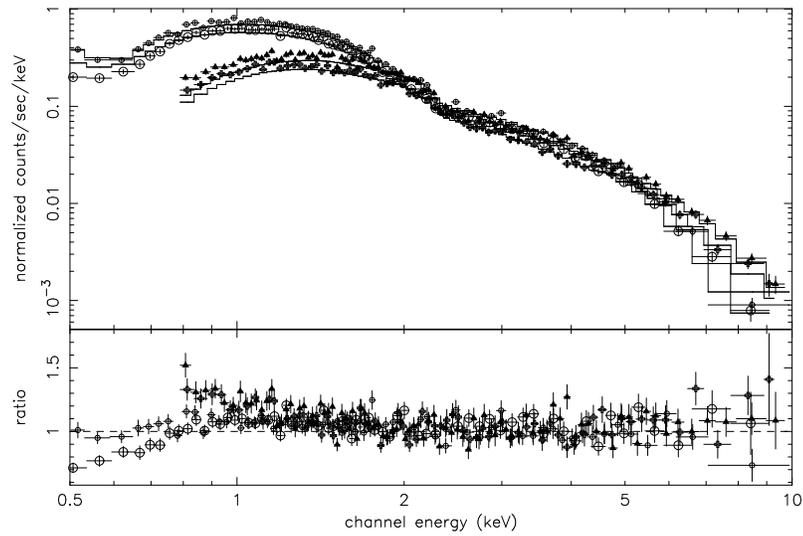}{7.cm}{270}{45}{45}{-200}{250}
\figcaption[fig1.ps]{
The ASCA spectra of PKS 0558-504 (upper panel) and the ratios of data to
a best power-law fit to the 2-10 keV band data (low panel). The SIS0, SIS1
and GIS2 and GIS3 data are marked with small circles, large circles,
crosses and triangles, respectively.  Systematic
excesses over the power-law are clearly present. The SIS spectrum is plotted
to 0.5 keV for the purpose of illustrating the degeneration of the SIS
efficiency.
\label{fig-1}}
\end{figure}

\begin{figure}
\figurenum{2} 
\epsscale{1.50}
\plotfiddle{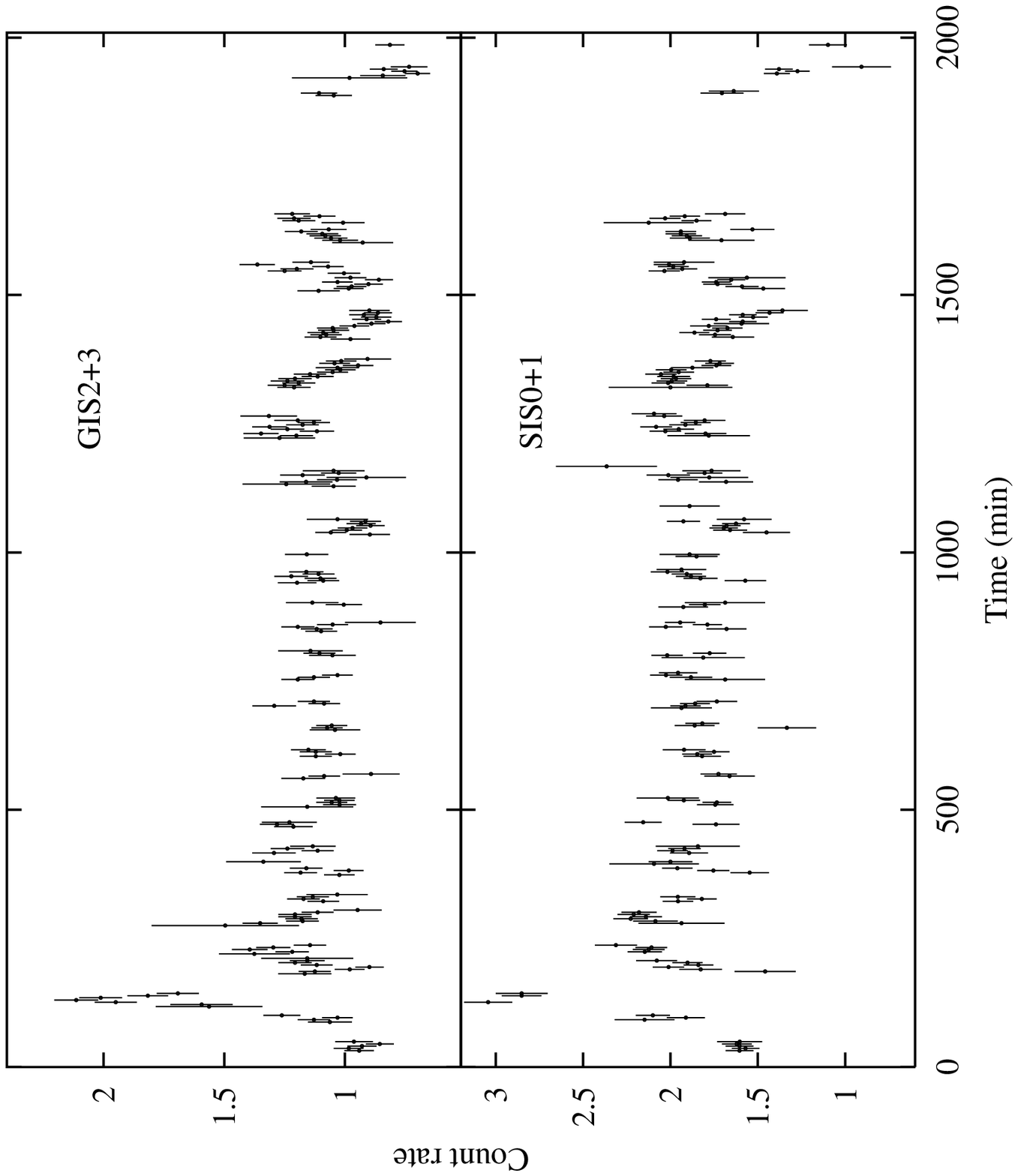}{12.cm}{270}{45}{45}{-200}{250}
\vspace{1cm}
\plotfiddle{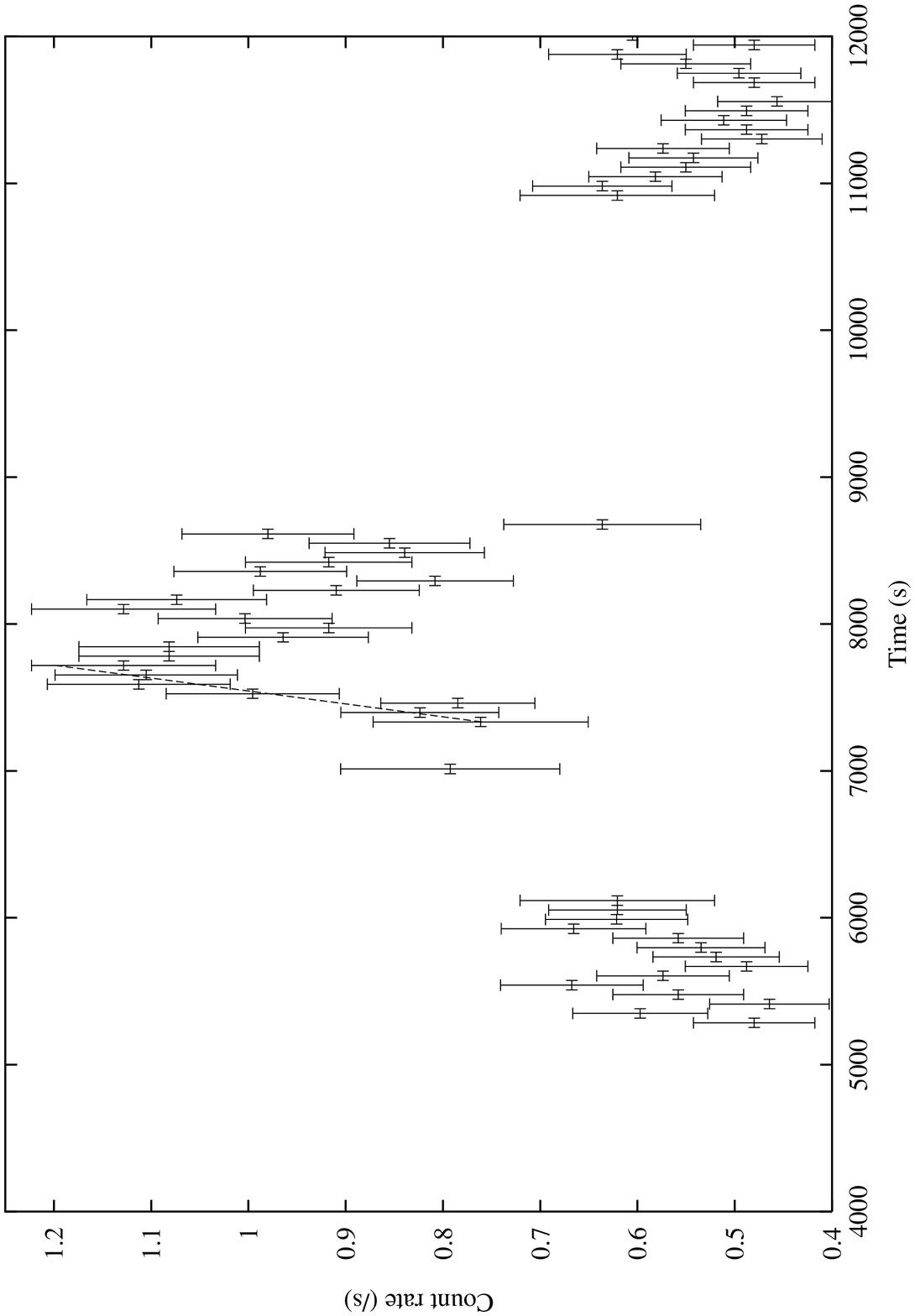}{6.cm}{270}{45}{45}{-200}{250}
\figcaption[fig2a.ps, fig2b.ps]{ The ASCA SIS and GIS light
curves for the PKS 0558-504 (Fig. 2a). The count rate is binned
in 256 s bins, and the data from two SIS and two GIS detectors
have been combined. The background count rate, which is about 5\%
of the total, has not been subtracted. The flare is clearly
present in the both SIS and GIS data, and is shown in enlarged
scale in the Fig. 2b for GIS data. A dashed line represents the
linear fit of the steep rising part the light curve for
estimation of the rate of change in the count rate. \label{fig-2}}
\end{figure}

\begin{figure}
\figurenum{3}
\epsscale{1.00}
\plotfiddle{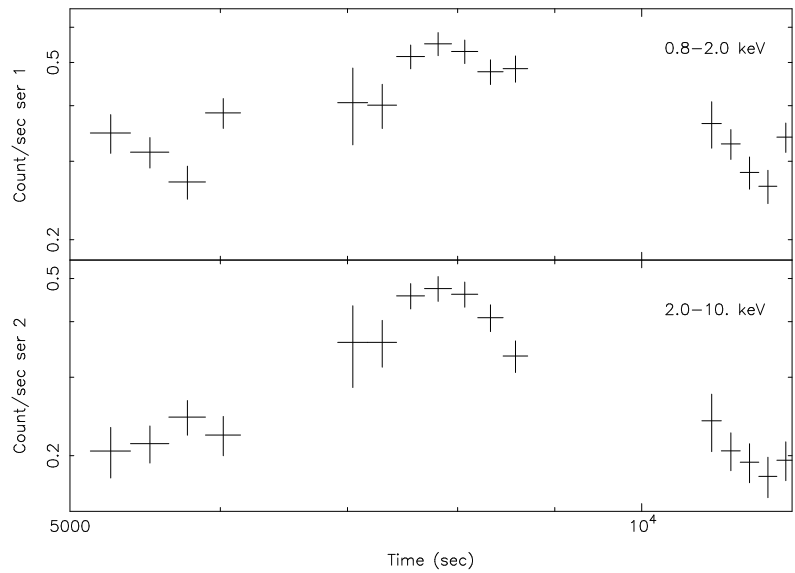}{12.cm}{270}{45}{45}{-200}{250}
\figcaption[fig3.ps]{
The hard and soft X-ray GIS light curves for the time period from
pre-flare to post-flare. Notice that the soft band shows different flare
profile with the hard one. The SIS curves show a similar character (not
shown here).
\label{fig-3}}
\end{figure}
\begin{figure}
\figurenum{4}
\epsscale{1.00}
\plotfiddle{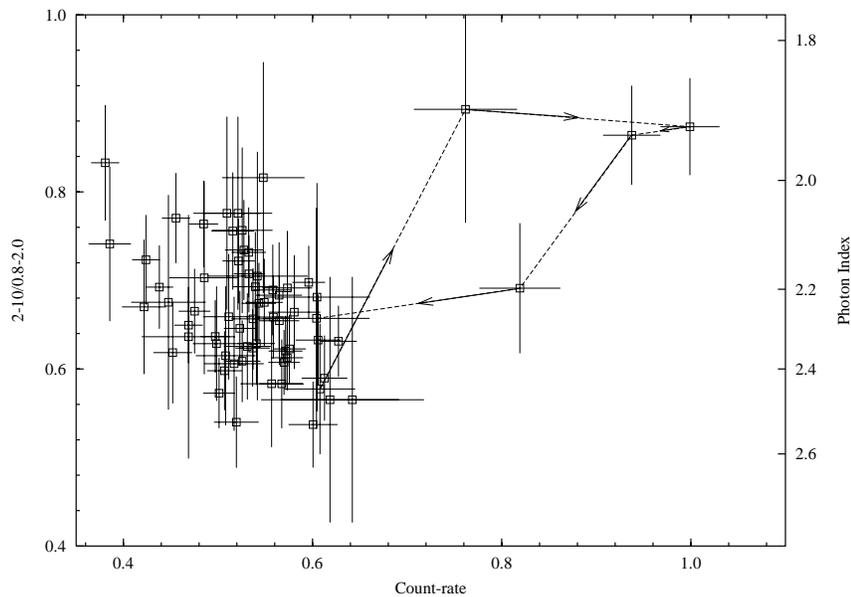}{6.cm}{270}{45}{45}{-200}{250}
\figcaption[fig4.ps]{
The plot of hardness ratio against the total count rate in the 0.8-10 keV
band for the SIS data. The right y-axis also shows the corresponding
photon index assuming a power-law spectrum.  The data points during the flare
and in the neighbor are connected with line and the arrows indicate the
proceeding direction of the flare. Notice that the flare show a clockwise
loop structure on the plot.
\label{fig-4}}
\end{figure}

\begin{thebibliography}{99}
\bibitem[Barr \& Mushotzky 1986]{bar86} Barr, P. \& Mushotzky, R.F.
1986, Nature, 320, 421
\bibitem[Beloborodov 1999]{bel99}Beloborodov, A.M., 1999, ApJ, 510, L123
\bibitem[Blandford \& Znajek 1977]{blan77}Blandford R.D., \& Znajek R.L., 1977, MNRAS, 179, 433
\bibitem[Boller, Brandt \& Fink 1996]{bol96}Boller, Th., Brandt, W.N., \&
Fink H.H. 1996, A\&A, 305, 53
\bibitem[Brinkmann, Yuan \& Siebert 1997]{bri97}Brinkmann W., Yuan W, Siebert J., 1997, A\&A, 319, 413
\bibitem[Catanese \& Sambruna 2000]{cat00} Catanese, M., \& Sambruna,
R.M. 2000, ApJ, 534, L39
\bibitem[Corbin 1997]{cor97}Corbin, M.R. 1997, ApJS, 113,245
\bibitem[Corbin & Smith 2000]{cor00}Corbin, M.R. 2000, ApJ, 532, 136
\bibitem[Fabian 1979]{fab79} Fabian, A.C. 1979, Proc. Roy. Soc. London A,
366, 449
\bibitem[Feigelsoni et al. 1986]{fei86}Feigelson, E., et al. 1986, ApJ, 302, 337
\bibitem[Ferrarese \& Merritt 2000]{fer00}Ferrarese L. \& Merritt D. 2000, ApJ, 539, L9
\bibitem[Fender 2000]{fen00}Fender, R.P., 2000, MNRAS, in press (astro-ph/0008447)
\bibitem[Forster \& Halpern 1996]{for96}Forster K., \& Halpern, J.P. 1996,
ApJ 468, 565
\bibitem[Galeev, Rosner \& Vaiana 1979]{grv79}Galeev, A.A., Rosner, R. \& Vaiana,G.S., 1979, ApJ, 229, 318
\bibitem[Gebhardt et al. 2000]{geb00}Gebhardt, K., et al., 2000, ApJ, 539, 13
\bibitem[Giommi et al. 1999]{gio99}Giommi, P., et al., 1999, A\&A, 351, 59
\bibitem[Gliozzi et al. 2000]{gli00}Gliozzi, M., Boller, Th., Brinkmann
W., Brandt, W.N. 2000, A\&A, 356,L17
\bibitem[Gliozzi et al. 2001]{gli01}Gliozzi, M., Brinkmann, W., O'Brien, P.T.,
Reeves, J.N., Pounds, K.A., Trifoglio, M. \& Gianotti, F. 2001, A\&A, 365, L128 
\bibitem[Goodrich 1989]{goo89} Goodrich R.W. 1989, ApJ, 342, 224
\bibitem[Gotthelf et al. 2000]{go00}Gotthelf E.V., Ueda Y., Fujimoto R., Kii T. \& Yamaoka K., 2000, ApJ, 543, 417
\bibitem[Guilbert, Fabian \& Rees 1983]{gui83} Guilbert P.W., Fabian,
A.C., \& Rees, M.J. 1983, MNRAS, 205, 593
\bibitem[Gregory et al. 1994]{Gre94}Gregory, P.C., Varasour, J.D., Scott, W.K., \& Condon, J.J. 1994,
ApJS, 90, 173
\bibitem[Grupe et al. 2000]{gru00}Grupe D., Leighly K.M., Thomas H.C., Laurent-Muehleisen S.A. 2000, A\&A, 356, 11
\bibitem[Kaspi et al. 2000]{kas00} Kaspi, S., Smith, P. S., Netzer, H., Maoz, D., Jannuzi, B. T., Giveon, U., 2000, ApJ, 533, 631
\bibitem[Kataoka et al. 2000]{kat00} Kataoka, J., Takahashi T., Makino
F., Inoue, S., Madejski G.M., Tashiro, M., Urry C.M., Kubo, H. 2000, ApJ, 528,
243
\bibitem[Kubo et al. 1998]{kub98}Kubo H., et al., 1998, ApJ, 504, 693
\bibitem[Leighly 1999a]{lei99a}Leighly, K., 1999a, ApJS, 125, 297
\bibitem[Leighly 1999b]{lei99b}Leighly, K., 1999b, ApJS, 125, 317
\bibitem[Laor 2000]{lao00}Laor A., 2000, ApJ, 543, L111
\bibitem[Magorrian et al. 1998]{mag98}Magorrian J., et al. 1998, AJ, 115, 2285
\bibitem[McHardy 1989]{mch89}McHardy, I. M. 1989, Proc., 23rd ESLAB
Symp. eds. J. Hunt \& B. Battrick (Paris:ESA) p. 499
\bibitem[McLure et al. 1999]{mc99}McLure, R. J., Kukula, M. J., Dunlop, J. S., Baum, S.
 A., O'Dea, C. P., Hughes, D. H.1999, MNRAS, 308, 377
\bibitem[Mineshige et al. 2000]{min00}Mineshige,S., Kawagichi, T., Takeuchi,
M., \& Hayashida, K., 2000, PASJ, 52, 499 
\bibitem[Moderski, Sikora \& Lasota 1998]{mod98}Moderski R., Sikora M., \& Lasota J.P. 1998, MNRAS, 301, 142
\bibitem[Nandra et al., 1997]{nan97} Nandra K. et al. 1997, ApJ, 476, 70
\bibitem[Nelson \& Whittle 1996]{nw96}Nelson C.H., Whittle, M., 1996, ApJ, 465, 96
\bibitem[Nelson 2000]{nel00}Nelson C.H., 2000, ApJ, 544, L91 
\bibitem[O'Brien et al., 2001]{obr01}O'Brien, P.T., et al., 2001, A\&A, 365, L122
\bibitem[Osterbrock \& Page 1985]{ost85}Osterbrock, D.E., \& Pagge R.W.
1985,  ApJ, 297, 166
\bibitem[Otani, Kii \& Miya 1996]{ota96} Otani, C., Kii, T., \& Miya, K.
1996, in R\"ontgenstrahlung from the Universe, ed. H.U. Zimmermann,
J.E. Tr\"umper \& H. Yorke (MPE Press, Garching), p. 491
\bibitem[Rees et al. 1982]{rees82}Rees M.J., Begelman M.C., Blandfor R.D., \& Phiney E.S. 1982, Nature, 295, 17
\bibitem[Rees 1984]{rees84} Rees, M.J. 1984, Ann. Rev. Astron. \&
Astrophy., 22, 471
\bibitem[Remillard et al. 1992]{rem92} Remillard, R.A., Grossan, B., Bradt
H.V., Ohashi T., Hayashida K., Makino F., \& Tanaka Y. 1992, Nature, 350,
589
\bibitem[Siebert et al. 1999]{sie99}Siebert J., Leighly K.M., Laurent-Muehleisen S.A., Brinkmann W., Boller Th. \& Matsuoka M. 1999, A\&A, 348, 678
\bibitem[Worrall \& Wilkes 1990]{wor90} Worrall, D.M., \& Wilkes, B.J.
1990, ApJ, 360, 396
\bibitem[Wright \& Otrupcek 1990]{wri90} Wright, A., \& Otrupcek, R., 1990,
Parkes Catalogue, Australia Telescope Nation Facility.
\bibitem[Wright et al. 1994]{wri94}Wright, A.E., Griffith, M.R., Burker, B.F.,
\& Ekers 1994, ApJS, 91, 111

\end{thebibliography}
\end{document}